\documentclass[prd,twocolumn]{revtex4}
\usepackage{graphicx, epsfig}
\usepackage{color}
\usepackage{mathrsfs}
\usepackage{amsmath}

\newcommand{\be}{\begin{equation}}
\newcommand{\ee}{\end{equation}}
\newcommand{\bea}{\begin{eqnarray}}
\newcommand{\eea}{\end{eqnarray}}

\newcommand{\gapp}{\mathrel{\raise.3ex\hbox{$>$}\mkern-14mu
              \lower0.6ex\hbox{$\sim$}}}
\newcommand{\lapp}{\mathrel{\raise.3ex\hbox{$<$}\mkern-14mu
              \lower0.6ex\hbox{$\sim$}}}

\begin{document}

\title{Hawking radiation as seen by an infalling observer}
\author{Eric Greenwood}
\author{Dejan Stojkovic}
\affiliation{HEPCOS, Department of Physics,
SUNY at Buffalo, Buffalo, NY 14260-1500}
\begin{abstract}
We investigate an important question of Hawking-like radiation as seen by an infalling observer during gravitational collapse.
Using the functional Schrodinger formalism we are able to probe the time dependent regime which is out of the reach of the standard approximations like the Bogolyubov method. We calculate the occupation number of particles whose frequencies are measured in the proper time of an infalling observer  in two crucially different space-time foliations: Schwarzschild and Eddington-Finkelstein. We demonstrate that the distribution in Schwarzschild reference frame is not quite thermal, though it becomes thermal once the horizon is crossed. We approximately fit the temperature and find that the local temperature increases as the horizon is approached, and diverges exactly at the horizon. In Eddington-Finkelstein reference frame the temperature at the horizon is finite, since the observer in that frame is not accelerated. These results are in agreement with what is generically expected in the absence of backreaction. We also discuss some subtleties related to the physical interpretation of the infinite local temperature in Schwarzschild reference frame.
\end{abstract}


\maketitle

\section{Introduction}
Understanding quantum radiation from collapsing objects, despite tremendous progress made in the last thirty years, still represents a major challenge in theoretical physics. An asymptotic observer, watching gravitational collapse of a massive object, will start registering radiation of quanta coming from the fields excited by the non-trivial metric of the background space-time. As the radiation progresses, it acquires more and more thermal features. Finally, when the horizon is formed, radiation becomes completely thermal, in agreement with the fact that radiation from a pre-existing horizon must be thermal, as shown for the first time by Hawking \cite{Hawking:1974sw}. In this picture, radiation is treated in the context of quantum mechanics, while the background is described by classical general relativity. Moreover, the background is held fixed, i.e. the form of the metric and even the mass of the collapsing object (or a black hole) are held fixed. It is extremely difficult to include any form of backreaction in the realistic models (see \cite{Candelas:1980zt,Russo:1992ax,Vachaspati:2007ur,Mottola:2006ew,
Mikovic:1997ty,Mikovic:1996bh} for some attempts).


In the absence of any backreaction, the radiation flux as seen by an asymptotic observer is constant at late times. If integrated over the infinite time (an asymptotic observer sees formation of a horizon only after infinite amount of his time) this flux diverges.
However, in classical general relativity, an infalling observer (say an observer which is falling together with a collapsing object) will cross the Schwarzschild radius in the finite amount of his proper time. This time interval corresponds to an infinite interval of the asymptotic observer. The question then is what such an observer would see. The answer will likely strongly depend on the space-time foliation one chooses. If an observer is infalling in Schwarzschild reference frame, his trajectory in any given moment corresponds to an accelerated observer. Though such an observer reaches the Schwarzschild radius in finite time, his acceleration diverges there and the temperature he measures will diverge. Such an observer would need to encounter all of the radiation that an asymptotic observer would see, only in finite time. Will such an observer be burned by an intensive radiation as he is sailing toward the Schwarzschild radius? Not necessarily. A "particle" has no invariant meaning. The distribution of produced particles (occupation number as a function of the frequency) strongly depends on the frequency of produced "particles". Since proper time coordinates for an asymptotic and infalling observers are different, it is likely that these two different observers will register different distributions of particles.

A very important complementary picture is given by Eddington-Finkelstein coordinates where the singularity at the horizon is absent. Such an observer would still register radiation due to the time dependent metric, however we do not expect that such radiation will be thermal in the whole frequency range. In addition, since the Schwarzschild radius is not a singular point, we do not expect the temperature of such radiation to diverge there.

Unfortunately, arguments of this type have remained only qualitative so far. The reason is that the notion of particles is well defined only in the asymptotically flat regions (e.g Minkowski or Schwarzschild). In the Bogolyubov method, which is widely used in this context, one matches the coefficients between the two asymptotically flat spaces, i.e. Minkowski at the beginning and Schwarzschild at the end of the gravitational collapse. The mismatch of these two vacua gives the number of produced particles. What happens in between is beyond the scope of the Bogolyubov method. Thus, the question what an infalling observer (in different space-time foliations) would register during the collapse can not be answered in the context of the Bogolyubov method.
However, recently developed functional Schrodinger formalism, goes beyond the approximations of the Bogolyubov method. In this context, one can successfully attack the question of radiation as seen by an infalling observer, which is the goal of this paper.

\section{Radiation as seen by an infalling observer in Schwarzschild reference frame}

We consider a spherical domain wall representing a spherical shell of collapsing matter. The wall is described by only the radial degree of freedom, $R(t)$. The metric is taken to be the solution of Einstein equations for a spherical domain wall.
The metric is Schwarzschild outside the wall, as follows
from spherical symmetry \cite{Ipser:1983db}
\begin{equation}
ds^2= -(1-\frac{R_s}{r}) dt^2 + (1-\frac{R_s}{r})^{-1} dr^2 +
      r^2 d\Omega^2 \ , \ \ r > R(t)
\label{metricexterior}
\end{equation}
where $R_s = 2GM$ is the Schwarzschild radius in terms of the mass,
$M$, of the wall, and
\begin{equation}
d\Omega^2  = d\theta^2  + \sin^2\theta d\phi^2 \, .
\end{equation}
In the interior of the spherical domain wall, the line element
is flat, as expected by Birkhoff's theorem,
\begin{equation}
ds^2= -dT^2 +  dr^2 + r^2 d\theta^2  + r^2 \sin^2\theta d\phi^2  \ ,
\ \ r < R(t)
\label{metricinterior}
\end{equation}
The equation of the wall is $r=R(t)$.
The interior time coordinate, $T$, is related to the asymptotic observer time coordinate, $t$, via the proper time of an observer moving with the shell, $\tau$. The relations are
\begin{equation}
\frac{dT}{d\tau} =
      \left [ 1 + \left (\frac{dR}{d\tau} \right )^2 \right ]^{1/2}
\label{bigTandtau}
\end{equation}
and
\begin{equation}
\frac{dt}{d\tau} = \frac{1}{B} \left [ B +
         \left ( \frac{dR}{d\tau} \right )^2 \right ]^{1/2}
\label{littletandtau}
\end{equation}
where
\begin{equation}
B \equiv 1 - \frac{R_s}{R}
\label{BofR}
\end{equation}
We will single out an observer who is falling together
with the collapsing shell, which means that he moves on
the trajectory $r = R(\tau)$ and measures the proper time $\tau$.
By integrating the equations of motion for the spherical domain wall, Ipser and Sikivie \cite{Ipser:1983db} found that the mass is a constant of motion and is given by
\begin{equation}
M = \frac{1}{2} [ \sqrt{1+R_\tau^2} + \sqrt{B+ R_\tau^2} ] 4\pi \sigma R^2
\label{ISmass}
\end{equation}
where $R_{\tau} = dR/d\tau$, while $\sigma$ is the surface tension (energy density per unit area) of the wall. It is assumed that ${\rm max} (R) < (4\pi G\sigma )^{-1} $ to avoid the case in which the domain wall is already within its own Schwarzschild radius to begin with. This expression for $M$ is implicit since $R_s =2GM$ occurs in $B$.
Solving for $M$ explicitly in terms of $R_\tau$ gives
\begin{equation}
M = 4\pi \sigma R^2 [ \sqrt{1+R_\tau^2} - 2\pi G\sigma R] .
\label{MRtau}
\end{equation}

From the mass, the velocity is given by
\begin{equation}
  |R_{\tau}|=\sqrt{\left(\frac{M}{4\pi\sigma R^2}+2\pi\sigma GR\right)^2-1}.
  \label{R_tau}
\end{equation}

Now consider a massless scalar field $\Phi$ which propagates in the background of the collapsing shell. The action for the scalar field is
\begin{equation} \label{S}
  S=\int d^4x\sqrt{-g}\frac{1}{2}g^{\mu\nu}\partial_{\mu}\Phi\partial_{\nu}\Phi \, ,
\end{equation}
where $g_{\mu \nu}$ is the background metric given by Eq. (\ref{metricexterior}) and Eq. (\ref{metricinterior}).
Decomposing the (spherically symmetric) scalar field into a complete set of real basis functions denoted by $\{f_k(r)\}$
\begin{equation}
  \Phi=\sum_ka_k(\tau)f_k(r)
  \label{mode expansion}
\end{equation}
we can find a complete set of independent eigenmodes $\{b_k\}$ for which the Hamiltonian is a sum of terms. The total wavefunction then factorizes and can be found by solving a time-dependent Schr\"{o}dinger equation of just one variable.

Since the metric inside and outside the shell have different forms, we can split the action (\ref{S}) into two parts

\begin{eqnarray}
  S_{in} &&= 2\pi\int d\tau\int_0^{R(\tau)}drr^2\left[-\frac{1}{\sqrt{1+R_{\tau}^2}}(\partial_{\tau}\Phi)^2 \right. \nonumber \\ && \left. +\sqrt{1+R_{\tau}^2}(\partial_r\Phi)^2\right]
 \end{eqnarray}

 \begin{eqnarray}
  S_{out} && =2\pi\int d\tau\int_{R(\tau)}^{\infty}drr^2\left[-\frac{B}{\sqrt{B+R_{\tau}^2}}\frac{(\partial_{\tau}\Phi)^2}{1-R_s/r} \right. \nonumber \\
  && \left.  + \frac{\sqrt{B+R_{\tau}^2}}{B}\left(1-\frac{R_s}{r}\right)(\partial_r\Phi)^2\right].
  \label{full action}
\end{eqnarray}
where we used (\ref{bigTandtau}) and (\ref{littletandtau}).

The most interesting things happen when the shell (and the infalling observer sitting on the shell) approaches the Schwarzschild radius.
From Eq.~(\ref{R_tau}) we see that $R_{\tau}$ is constant in the limit when $R\rightarrow R_s$.
Therefore the kinetic term for $S_{in}$ is roughly constant. The kinetic term in $S_{out}$ goes to zero as $R\rightarrow R_s$, so the $S_{in}$ kinetic term is dominant. Similarly the potential term in $S_{in}$ goes to a constant while the potential term in $S_{out}$ becomes very large, so the potential term in $S_{out}$ dominates.
Strictly speaking, this argument fails in the neighborhood of  $r = R$, however the dominant contribution to the integrals will be for $r \neq R$. Therefore we can write the action as
\begin{eqnarray}
  S \approx 2\pi && \int d\tau\left[-\int_{0}^{R_s}drr^2\frac{1}{\sqrt{1+R_{\tau}^2}}(\partial_{\tau}\Phi)^2 \right. \nonumber \\
  &&\left.+\int_{R_s}^{\infty}drr^2\frac{|R_{\tau}|}{B}\left(1-\frac{R_s}{r}\right)(\partial_r\Phi)^2\right]
          \label{Action}
\end{eqnarray}
where we have changed the limits of integration from $R(\tau)$ to $R_s$ since this is the region of interest.

Using the expansion in the modes Eq.~(\ref{mode expansion}), we can rewrite the action as
\begin{eqnarray}
  S \approx && \int d\tau\left[-\frac{1}{2}\frac{1}{\sqrt{1+R_{\tau}^2}}\dot{a}_k(\tau)\mathbf{A}_{kk'}\dot{a}_{k'}(\tau) \right. \nonumber \\
  && \left. +\frac{|R_{\tau}|}{2B}a_k(\tau)\mathbf{C}_{kk'}a_{k'}(\tau)\right]
\end{eqnarray}
where $\dot{a}=da/d\tau$, and $\mathbf{A}$ and $\mathbf{C}$ are matrices that are independent of $R(\tau)$ and are given by
\begin{align}
  \mathbf{A}_{kk'}&=4\pi\int_{0}^{R_s}drr^2f_k(r)f_{k'}(r)\\
  \mathbf{C}_{kk'}&=4\pi\int_{R_s}^{\infty}drr^2\left(1-\frac{R_s}{r}\right)f_k'(r)f_{k'}'(r).
\end{align}

From the action
(\ref{Action}) we can find the Hamiltonian, and according to the standard quantization procedure, the wave function $\psi(a_k,\tau)$ must satisfy
\be
i\frac{\partial\psi}{\partial\tau} =H \psi \, ,
\ee
or
\begin{eqnarray}
&& i\frac{\partial\psi}{\partial\tau} =  \\ &&  \left[\frac{1}{2}\sqrt{1+R_{\tau}^2}\Pi_k(\mathbf{A}^{-1})_{kk'}\Pi_{k'}+ \frac{|R_{\tau}|}{2B}a_k(\tau)\mathbf{C}_{kk'}a_{k'}(\tau)\right]\psi \nonumber
\end{eqnarray}
where
\begin{equation}
  \Pi_k=-i\frac{\partial}{\partial a_k(\tau)}
\end{equation}
is the momentum operator conjugate to $a_k(\tau)$.

So the problem of radiation from the collapsing domain wall for the infalling observer is equivalent to the problem of an infinite set of coupled harmonic oscillators with time dependent frequency. Since $\mathbf{A}$ and $\mathbf{C}$ are symmetric and real (i.e. Hermitian), it is possible to simultaneously diagonalize them using the principal axis transformation. Then for a single eigenmode, the Schr\"{o}dinger equation takes the form
\begin{equation}
  \left[-\frac{1}{2m}\sqrt{1+R_{\tau}^2}\frac{\partial^2}{\partial b^2}+\frac{|R_{\tau}|}{2B}Kb^2\right]\psi(b,\tau)=i\frac{\partial\psi(b,\tau)}{\partial\tau}
  \label{b Shrod}
\end{equation}
where $m$ and $K$ denote eigenvalues of $\mathbf{A}$ and $\mathbf{C}$, and $b$ is the eigenmode.

Re-writing Eq.~(\ref{b Shrod}) in the standard form we obtain
\begin{equation}
  \left[-\frac{1}{2m}\frac{\partial^2}{\partial b^2}+\frac{m}{2}\omega^2(\eta)b^2\right]\psi(b,\eta)=i\frac{\partial\psi(b,\eta)}{\partial\eta}
  \label{new b Shrod}
\end{equation}
where
\begin{equation} \label{omega_sq}
  \omega^2(\eta)=\frac{K}{m}\frac{|R_{\tau}|}{B\sqrt{1+R_{\tau}^2}}\equiv\omega_0^2\frac{|R_{\tau}|}{B\sqrt{1+R_{\tau}^2}}
\end{equation}
and
\begin{equation}
  \eta=\int d\tau'\sqrt{1+R_{\tau}^2}.
  \label{Eta}
\end{equation}
where we defined $\omega_0^2 \equiv K/m$.
The exact solution to Eq.~(\ref{new b Shrod}) is given by \cite{Dantas}
\begin{equation}
  \psi(b,\eta)=e^{i\alpha(\eta)}\left(\frac{m}{\pi\rho^2}\right)^{1/4}\exp\left[\frac{im}{2}\left(\frac{\rho_{\eta}}{\rho}+\frac{i}{\rho^2}\right)b^2\right]
  \label{wavfunc}
\end{equation}
where $\rho_{\eta}=d\rho/d\eta$ and $\rho$ is given by the real solution of the ordinary (though non-linear) differential equation
\begin{equation}
  \rho_{\eta\eta}+\omega^2(\eta)\rho=\frac{1}{\rho^3}
  \label{rho eq}
\end{equation}
with initial conditions
\begin{equation}
  \rho(0)=\frac{1}{\sqrt{\omega_0}},\hspace{3mm} \rho_{\eta}(0)=0.
  \label{rho IC}
\end{equation}
The phase $\alpha$ is given by
\begin{equation}
  \alpha(\eta)=-\frac{1}{2}\int_0^{\eta}\frac{d\eta'}{\rho^2(\eta')}.
  \label{alpha}
\end{equation}
Complete information about the radiation in the background of the collapsing shell is contained in the wavefunction (\ref{wavfunc}).

Consider an observer with detectors that are designed to register particles of different frequencies for the free field $\Phi$. Such an observer will interpret the wavefunction of a given mode $b$ at some later time in terms of simple harmonic oscillator states, $\{\varphi_n\}$, at the final frequency, $\bar{\omega}$. The initial ($\tau=0$) vacuum state for the modes is the simple harmonic oscillator ground state
\be
\varphi (b) = \left(\frac{m\omega_0}{\pi} \right)^{1/4} e^{-m\omega_0 b^2/2} \, .
\ee
The relationship between the coordinates $\eta$ and $\tau$ is given by Eq.~(\ref{Eta}), which near $R_s$ simplifies to $d\eta / d\tau=$const.
The number of quanta in eigenmode $b$ can be evaluated by decomposing Eq.~(\ref{wavfunc}) in terms of the states and evaluating the occupation number of that mode. The wavefunction for a given mode in terms of simple harmonic oscillator basis  is given by
\begin{equation}
   \psi(b,\tau)=\sum_nc_n(\tau)\varphi(b)
\end{equation}
where
\begin{equation}
  c_n=\int db\varphi_n^*(b)\psi(b,v)
  \label{c num}
\end{equation}
which is an overlap of the wavefunction at some later time $\psi(b,\tau)$ with the simple harmonic oscillator basis functions. The occupation number at eigenfrequency $\bar{\omega}$ is given by
\begin{equation}
  N(\tau,\bar{\omega})=\sum_n n |c_n|^2.
\end{equation}
The occupation number in the eigenmode $b$ is then given by (see Appendix A)
\begin{equation} \label{N}
  N(\tau,\bar{\omega})=\frac{\bar{\omega}\rho^2}{\sqrt{2}}\left[\left(1-\frac{1}{\bar{\omega}\rho^2}\right)^2+
  \left(\frac{\rho_{\tau}}{\bar{\omega}\rho}\right)^2\right] \, .
\end{equation}
where $\rho_{\tau}=d\rho / d\tau$.

\begin{figure}[htbp]
\centering
\includegraphics{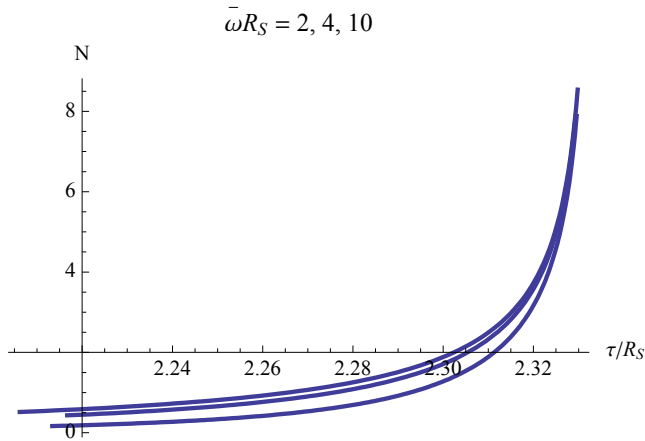}
\caption{The occupation number $N$ as a function of the proper time $\tau/R_s$ for various fixed values of particle frequencies $\bar{\omega}R_s$. The curves are lower for higher values of $\bar{\omega}R_s$. The occupation number diverges as the infalling observer approaches $R_s$, which happens as $\tau \rightarrow \tau_c = 2.33333... R_s^{-1}$.}
\label{Nvtau}
\end{figure}

For fixed $\bar{\omega}$, $N$ is a function of time since $\rho $ and $\rho_{\tau}$ are functions of time.
In Fig.~\ref{Nvtau} we plot the occupation number of produced particles as a function of time (for several fixed frequencies $\bar{\omega}R_s$). The amount of proper time needed for the shell (and the infalling observer) to reach $R_s$ can be obtained by integrating Eq.~(\ref{R_tau}). For $\sigma = 0.01 R_s^{-3}$ this critical proper time is $\tau_c = 2.33333... R_s^{-1}$. Fig.~\ref{Nvtau} shows that, as the infalling observer approaches $R_s$, the occupation number increases and diverges exactly at $R_s$. The same conclusion can be drawn by analyzing the occupation number $N$ in Eq.~(\ref{N}) as a function of $\rho$ and $\rho_\tau$ (see Appendix B). This is in agreement with what one would expect in the absence of backreaction. Hawking showed in \cite{Hawking:1974sw} that, for an asymptotic observer, the flux of particles at late times is steady (constant in time). This means that it diverges for a fixed background (i.e. fixed mass of the collapsing object) since we effectively keep pumping energy into the collapsing object so that its mass is constant despite radiation. For the asymptotic observer it takes infinite amount of his time for the collapsing object to reach its own Schwarzschild radius.  This infinite interval corresponds to a finite time interval of the infalling observer's time. Thus, one may conclude that the infalling observer has to encounter an infinite number of particles before he hits $R_s$ \cite{BirrellandDavies}.
However, we should keep in mind that it is the particle occupation number that is divergent, not the actual number of particles detected by the infalling observer. The number of particles detected may be quite different from the occupation number. As the infalling observer approaches $R_s$, he and his detector get blue-shifted.
His detector can not register particles whose wavelength is larger than his detector; for them the detector is "inside"
the particles \cite{BirrellandDavies}. Thus, an infalling observer might not be able to register all of the created particles.
We only calculated the occupation number of particles contained in the wave function describing radiation at some later time. The occupation number does not tell us where the particle are nor how they propagate further.

\begin{figure}[htbp]
\centering
\includegraphics{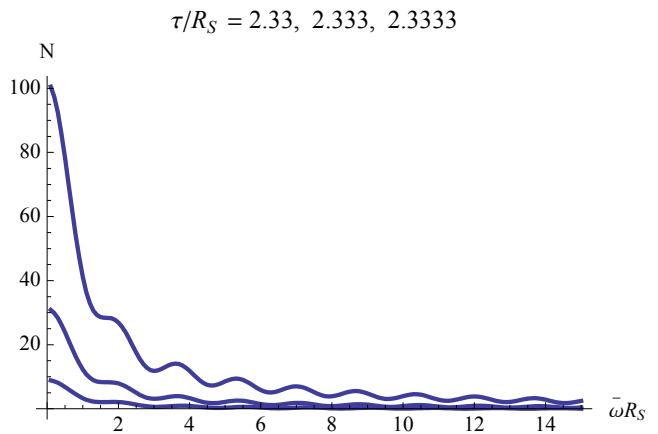}
\caption{
The occupation number $N$ as a function of frequency $\bar{\omega}R_s$ for fixed values of the proper time $\tau/R_s$. The occupation number increases for larger values of $\tau/R_s$ as $\tau \rightarrow \tau_c = 2.33333... R_s^{-1}$.}
\label{Nvomega}
\end{figure}

For fixed time $\tau$, the occupation number $N$ is a function of frequency $\bar{\omega}$ at some fixed $\tau$.
From Eq.~(\ref{omega_sq}) we see that, in order to keep the values of $\bar{\omega}$ fixed in time, $\omega_0\rightarrow0$ as $B \rightarrow 0$. Thus, $\bar{\omega}$ varies with  $\omega_0$ and not with time since all of the values for $\bar{\omega}$ must be calculated at the same final time.
In Fig.~\ref{Nvomega} we plot the occupation number $N$ as a function of frequency $\bar{\omega}R_s$ for fixed values of the proper time $\tau/R_s$. We can compare these plots with the occupation numbers for the thermal Planck distribution
\begin{equation}
  N_{\rm Planck} (\bar{\omega})=\frac{1}{e^{\beta \bar{\omega}}-1},
  \label{Planck Dist}
\end{equation}
where $\beta$ is the inverse temperature. The curves have manifest non-thermal features --- the occupation number does not diverge at $\bar{\omega}R_s = 0$ as in the Planck distribution and small scale oscillations which were absent in the Planck distribution are present here.  However, as $\tau\rightarrow\tau_c = 2.33333... R_s^{-1}$ the distribution becomes more thermal.

\begin{figure}[htbp]
\centering
\includegraphics{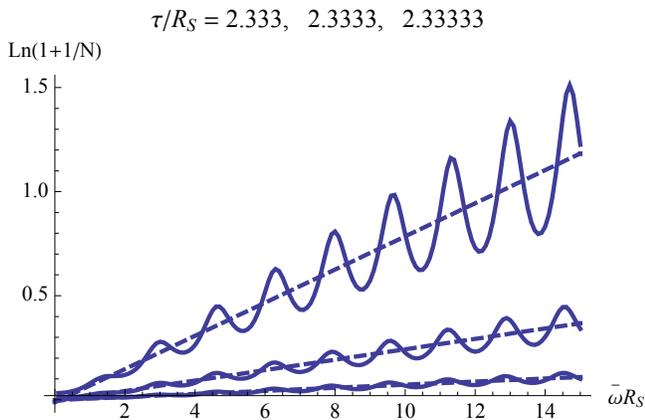}
\caption{Plot of $\ln(1+1/N)$ as a function of frequency $\bar{\omega}R_s$ for fixed values of the proper time $\tau/R_s$.  The slope of the best fit line is $\beta$, which is the inverse temperature. The non-thermal features disappear and the temperature diverges as the Schwarzschild radius is approached, i.e. $\tau \rightarrow \tau_c = 2.33333... R_s^{-1}$.}
\label{LNvom}
\end{figure}

In Fig.~\ref{LNvom} we fit the best fit linear approximation of the spectrum in Fig.~\ref{Nvomega}. From Eq.~(\ref{Planck Dist}) we see that the slope of the line is $\beta$, which is proportional to the inverse temperature, i.e. $T\propto\beta^{-1}$. Several important features of the Hawking-like radiation can be read off this plot. Non-thermal features of the radiation are apparent. It is interesting that, exactly as argued in \cite{VSK}, departures from thermality are larger for larger frequencies. As the time progresses and the infalling observer is approaching $R_s$, radiation becomes more and more thermal even at large frequencies. Finally, at $R_s$ it becomes purely thermal. At that point a black hole is formed and it is natural that the spectrum become thermal, as known from various studies of quantum radiation from a pre-existing horizon. We can not extrapolate the results for any further evolution of the infalling observer since the near-horizon approximation used in Eq.~(\ref{Action}) fails. In the standard picture, the infalling observer will reach the singularity, i.e. a region of infinitely strong gravitational field, in some finite proper time, and any results about produced particles in that regime are likely to be non-physical. This may not happen if the quantum effects are capable of erasing the classical singularity at the center, just as in atomic physics quantum mechanics got rid of the singularity of the Coulomb potential which has an identical 1/r behavior  \cite{Greenwood:2008ht}.

It is apparent that the slope of the  $\ln(1+1/N)$ versus $\bar{\omega}R_s$ curve is decreasing as the infalling observer is approaching $R_s$. Exactly at $R_s$, the slope of the curve is zero indicating that the temperature of the radiation is infinite. This is not surprising since, as it is well known, the asymptotic observer in the nearly flat asymptotic region will register Hawking radiation with some finite temperature. When this temperature is blue-shifted back to $R_s$, it clearly diverges. Thus, the occupation number of particles as seen by an infalling observer will have a distribution with the temperature which diverges as $R_s$ is approached.

Infinite near horizon local temperature of the Hawking-like radiation may or may not indicate that the backreaction due to radiation will be important in that region. The standard lore is that such backreaction is small. For that question, the relevant quantity which needs to be calculated is the stress-energy tensor near the horizon. As pointed out in \cite{Davies:1976ei}, where a simple $1+1$ dimensional model was studied, the local value of the stress-energy tensor is small since the local vacuum polarization cancels out the divergent temperature energy density due to radiation. However, one should keep in mind that a complete realistic $3+1$ dimensional calculation does not exist. Also, it may happen that a local backreaction is small but a global integrated backreaction can not be neglected since a collapsing object excites fields which are not localized in space and time. These are the questions of great importance in black hole physics but are outside of the scope of this work.

\section{Radiation in ingoing Eddington-Finkelstein coordinates}

Now we consider the collapse from the point of view of an ingoing Eddington-Finkelstein observer. This is a different space-time foliation than that in Schwarzschild coordinates, and we expect crucially different results. In particular, since the metric is not divergent at the horizon, we do not expect infinite temperature there.

For this purpose, we define the ingoing null coordinate $v$ as
\be
  v=t+r^*
\ee
where $r^*$ is the tortoise coordinate. We can then rewrite Eq.~(\ref{metricexterior}) as
\be
  ds^2=-\left(1-\frac{R_s}{r}\right)dv^2+2dvdr+r^2d\Omega^2, \hspace{2mm} r>R(v).
  \label{out_Met}
\ee
where the trajectory of the collapsing wall is $r=R(v)$.
The interior metric is the same as in Eq.~(\ref{metricinterior})
The interior time coordinate, $T$, is related to the ingoing null coordinate, $v$, via the proper time on the shell, $\tau$. [Note that the proper time $\tau$ is different from the same quantity in Schwarzschild coordinates since the space-time foliation is different.]  The relations are
\be
  \frac{dT}{d\tau}=\sqrt{1+\left(\frac{dR}{d\tau}\right)^2}
  \label{dTdtau}
\ee
and
\be
  \frac{dv}{d\tau}=\frac{1}{B}\left(\frac{dR}{d\tau}-\sqrt{B+\left(\frac{dR}{d\tau}\right)^2}\right)
  \label{dvdtau}
\ee
where
\be
  B\equiv 1-\frac{R_s}{R}.
\ee

Consider again a massless scalar field $\Phi$ which propagates in the background of the collapsing shell. The action for the scalar field is
\be
  S=\int d^4x\sqrt{-g}\frac{1}{2}g^{\mu\nu}\partial_{\mu}\Phi\partial_{\nu}\Phi,
  \label{action}
\ee
where $g^{\mu\nu}$ is the background metric given by Eqs.~(\ref{metricinterior}) and (\ref{out_Met}). Decomposing the (spherically symmetric) scalar field into a complete set of real basis functions denoted by $\{f_k(r)\}$
\be
  \Phi=\sum_ka_k(v)f_k(r)
  \label{mode_ex}
\ee
we can find a complete set of independent eigenmodes $\{b_k\}$ for which the Hamiltonian is a sum of terms.

Since the metric inside and outside of the shell have different forms, we again split the action Eq.~(\ref{action}) into two parts
\be
  S_{in}=2\pi \int dT\int_0^{R(v)}drr^2\left[-(\partial_T\Phi)^2+(\partial_r\Phi)^2\right],
  \label{S_in}
\ee
\begin{align}
  S_{out}=2\pi\int dv\int_{R(v)}^{\infty}drr^2&\Big{[}\partial_v\Phi\partial_r\Phi+\partial_r\Phi\partial_v\Phi\nonumber\\
  &+\left(1-\frac{R_s}{r}\right)(\partial_r\Phi)^2\Big{]}.\label{S_out}
\end{align}
We are again interested in the near horizon behavior of the radiation, i.e. as $R\rightarrow R_s$. In this limit we can write Eq.~(\ref{dvdtau}) as
\be \label{app}
  \frac{dv}{d\tau}\approx-\frac{1}{2R_{\tau}}
\ee
where $R_{\tau}=dR/d\tau$. The explicit functional dependence of $R_{\tau}$ can be found again from Eq.~(\ref{R_tau}), which is just the consequence of the fact that $M$ in Eq.~(\ref{ISmass}) is the conserved quantity also in Eddington-Finkelstein coordinates. Then with the help of Eq.~(\ref{dTdtau}) we can write Eq.~(\ref{S_in}) as
\begin{align}
  S_{in}=2\pi\int dv\int_0^{R(v)}drr^2&\Big{[}-\frac{1}{2}\frac{1}{\sqrt{R_v/2(R_v/2+1)}}(\partial_v\Phi)^2\nonumber\\
  &+2\sqrt{R_v/2(R_v/2+1)}(\partial_r\Phi)^2\Big{]}\label{S_in_v}
\end{align}
where $R_v=dR/dv$. Obviously, the action is not singular as $R(v) \rightarrow R_s$, unlike the Schwarzschild case. From Eqs.~(\ref{S_out}) and (\ref{S_in_v}) we can write the total action as
\begin{align}
  S\approx&2\pi\int dv\Big{[}-\int_0^{R_s}drr^2\frac{1}{2}\frac{1}{\sqrt{R_v/2(R_v/2+1)}}(\partial_v\Phi)^2\nonumber\\
  &+\int_{R_s}^{\infty}drr^2\partial_v\Phi\partial_r\Phi+\int_{R_s}^{\infty}drr^2\partial_r\Phi\partial_v\Phi\nonumber\\
  &+\int_{R_s}^{\infty}drr^2\left(1-\frac{R_s}{r}\right)(\partial_r\Phi)^2\Big{]}
\end{align}
where we have changed the limits of integration from $R(v)$ to $R_s$ since this is the region of interest.

Now using the expansion in modes Eq.~(\ref{mode_ex}), we can rewrite the action as
\begin{align}
  S\approx\int dv&\Big{[}-\frac{1}{2}\frac{1}{\sqrt{R_v/2(R_v/2+1)}}\dot{a}_k{\bf A}_{kk'}\dot{a}_{k'}\nonumber\\
  &+\frac{1}{2}\dot{a}_k{\bf Y}_{kk'}a_{k'}+\frac{1}{2}a_k{\bf Y}_{kk'}^{-1}\dot{a}_{k'}+\frac{1}{2}a_k{\bf C}_{kk'}a_{k'}\Big{]}
\end{align}
where $\dot{a}=da/dv$, and ${\bf A}$, ${\bf Y}$ and ${\bf C}$ are matrices that are independent of $R(v)$ and are given by
\begin{align}
  {\bf A}_{kk'}=2\pi\int_0^{R_s}drr^2f_k(r)f_{k'}(r),\\
  {\bf Y}_{kk'}=4\pi\int_{R_s}^{\infty}drr^2f_k(r)f'_{k'}(r),\\
  {\bf C}_{kk'}=8\pi\int_{R_s}^{\infty}drr^2\left(1-\frac{R_s}{r}\right)f'_k(r)f'_{k'}(r).
\end{align}
However if we take that the matrices are symmetric and real, we can see that ${\bf Y}={\bf Y}^{-1}$, so we can write the action as
\begin{align}
  S\approx\int dv&\Big{[}-\frac{1}{2}\frac{1}{\sqrt{R_v/2(R_v/2+1)}}\dot{a}_k{\bf A}_{kk'}\dot{a}_{k'}\nonumber\\
  &+\frac{1}{2}{\bf Y}_{kk'}\left(\dot{a}_ka_{k'}+a_k\dot{a}_{k'}\right)+\frac{1}{2}a_k{\bf C}_{kk'}a_{k'}\Big{]}.\label{Action}
\end{align}

From the action Eq.~(\ref{Action}) we can find the Hamiltonian, and according to the standard quantization procedure, the wave function $\psi(a_k,v)$ must satisfy
\be
  i\frac{\partial\psi}{\partial v}=H\psi,
\ee
or
\begin{align}
  i\frac{\partial\psi}{\partial v}=&\Big{[}\frac{1}{2}\sqrt{R_v/2(R_v/2+1)}\Pi_k({\bf A^{-1}})_{kk'}\Pi_{k'}\nonumber\\
  &+\frac{1}{2}a_k\left(\sqrt{R_v/2(R_v/2+1)}{\bf Y}^2_{kk'}({\bf A^{-1}})_{kk'}+{\bf C}_{kk'}\right)a_{k'}\nonumber\\
  &+\frac{1}{2}\sqrt{R_v/2(R_v/2+1)}\Pi_k{\bf Y}_{kk'}({\bf A^{-1}})_{kk'}a_{k'}\Big{]}\psi
\end{align}
where
\be
  \Pi_k=-i\frac{\partial}{\partial a_k}
\ee
is the momentum operator conjugate to $a_k$. Using the momentum $\Pi_k$, we can rewrite the Schr\"odinger equation as
\begin{align}
  i\frac{\partial\psi}{\partial v}=&\Big{[}\frac{1}{2}\sqrt{R_v/2(R_v/2+1)}\Pi_k({\bf A^{-1}})_{kk'}\Pi_{k'}\nonumber\\
  &+\frac{1}{2}a_k\left(\sqrt{R_v/2(R_v/2+1)}{\bf Y}^2_{kk'}({\bf A^{-1}})_{kk'}+{\bf C}_{kk'}\right)a_{k'}\nonumber\\
  &-i\frac{1}{2}\sqrt{R_v/2(R_v/2+1)}{\bf Y}_{kk'}({\bf A^{-1}})_{kk'}\delta_{kk'}\Big{]}\psi
\end{align}
where $\delta_{kk'}$ is the Kronecker delta function.

So the problem of radiation from the collapsing domain wall for the infalling observer is equivalent to the problem of an infinite set of decoupled damped harmonic oscillators with time-dependent frequency. Since ${\bf A}$, ${\bf Y}$ and ${\bf C}$ are symmetric and real, it is possible to simultaneously diagonalize them using the principle axis transformation. Then for a single eigenmode, the Schr\"odinger equation takes the form
 \begin{align}
  i\frac{\partial\psi}{\partial v}=&\Big{[}-\frac{1}{2m}\sqrt{R_v/2(R_v/2+1)}\frac{\partial^2}{\partial b^2}\nonumber\\
  &+\frac{1}{2}\left(\sqrt{R_v/2(R_v/2+1)}\frac{y^2}{m}+K\right)b^2\nonumber\\
  &-i\frac{y}{2m}\sqrt{R_v/2(R_v/2+1)}\Big{]}\psi
  \label{schrod}
\end{align}
where $m$, $y$ and $K$ denote eigenvalues of ${\bf A}$, ${\bf Y}$ and ${\bf C}$, and $b$ is the eigenmode.

Re-writing Eq.~(\ref{schrod}) in the standard form we obtain
\be
  \left[-\frac{1}{2m}\frac{\partial^2}{\partial b^2}+\frac{m}{2}\omega^2(\eta)-i\frac{y}{2m}\right]\psi(b,\eta)=i\frac{\partial\psi(b,\eta)}{\partial\eta}
  \label{Schrod}
\ee
where
\begin{align}
  \omega^2(\eta)&=\frac{y^2}{m^2}+\frac{K}{m}\frac{1}{\sqrt{R_v/2(R_v/2+1)}}\nonumber\\
        &\equiv\frac{y^2}{m^2}+\frac{\omega_0^2}{\sqrt{R_v/2(R_v/2+1)}}
\end{align}
and
\be
  \eta=\int dv'\sqrt{R_v/2(R_v/2+1)}
\ee
where we defined $\omega_0^2\equiv K/m$. To find solutions to equation Eq.~(\ref{Schrod}) we use the ansatz
\be
  \psi(b,\eta)=e^{-y\eta/2m}\phi(b,\eta).
  \label{ansatz}
\ee
This leads to the equation for $\phi(b,\eta)$
\be
  -\frac{1}{2m}\frac{\partial^2\phi}{\partial b^2}+\frac{m\omega^2}{2}b^2\phi=i\frac{\partial\phi}{\partial\eta}.
\ee
As discussed in Ref.~\cite{Dantas}, this has the implicit solution
\be
  \phi(b,\eta)=e^{i\alpha(\eta)}\left(\frac{m}{\pi\rho^2}\right)^{1/4}\exp\left[\frac{im}{2}\left(\frac{\rho_{\eta}}{\rho}+\frac{i}{\rho^2}\right)b^2\right]
  \label{phi}
\ee
where $\rho_{\eta}$ is the derivative of the function $\rho(\eta)$ with respect to $\eta$, and the defining equation for $\rho$ is
\be
  \rho_{\eta\eta}+\omega^2(\eta)\rho=\frac{1}{\rho^3}.
\ee
The initial conditions for $\rho$ are taken at some large value of $\eta$ (i.e. large value of $R$) denoted by $\eta_i$, so that
\be
  \rho(\eta_i)=\frac{1}{\sqrt{\omega(\eta_i)}}, \hspace{2mm} \rho_{\eta}(\eta_i)=0.
\ee
The phase $\alpha$ is defined by
\be
  \alpha(\eta)=-\frac{1}{2}\int^{\eta}\frac{d\eta'}{\rho^2(\eta')}.
\ee
Then Eq.~(\ref{ansatz}) give
\be
  \psi=e^{-y\eta/2m}\phi(b,\eta)
  \label{psi}
\ee
where $\phi$ given in Eq.~(\ref{phi}).

Consider an observer with detectors that are designed to register particles of different frequencies for the free field $\Phi$ at early times. Such an observer will interpret the wavefunction of a given mode $b$ at late times in terms of simple harmonic oscillator states, $\{\varphi_n\}$, at the final frequency $\bar{\omega}$. The number of quanta in eigenmode $b$ can be evaluated by decomposing the wavefunction Eq.~(\ref{psi}) in terms of the states, $\{\varphi_n\}$, and by evaluating the occupation number of that mode. To implement this evaluation, we start by writing the wavefunction for a given mode at time $v<v_f$ in terms of the simple harmonic oscillator basis at $v=v_0$
\be
  \psi(b,v)=\sum_nc_n(v)\varphi_n(b)
\ee
where
\be
  c_n=\int db\varphi_n^*(b)\psi(b,v)
\ee
which is the Gaussian overlap with the simple harmonic oscillator basis functions. The occupation number at eigenfrequency $\bar{\omega}$ by the time $v<v_f$, is given by the expectation value
\be
  N(v,\bar{\omega})=\sum_n\left|c_n\right|^2.
\ee
We evaluate the occupation number in the eigenmode $b$ to be
\be
  N(v,\bar{\omega})=\frac{\bar{\omega}\rho^2}{\sqrt{2}}e^{-y\eta/m}
  \left[\left(1-\frac{1}{\bar{\omega}\rho^2}\right)^2+\left(\frac{\rho_{\eta}}{\bar{\omega}\rho}\right)^2\right]
\ee
for $v<v_f$.

\begin{figure}[htbp]
\includegraphics{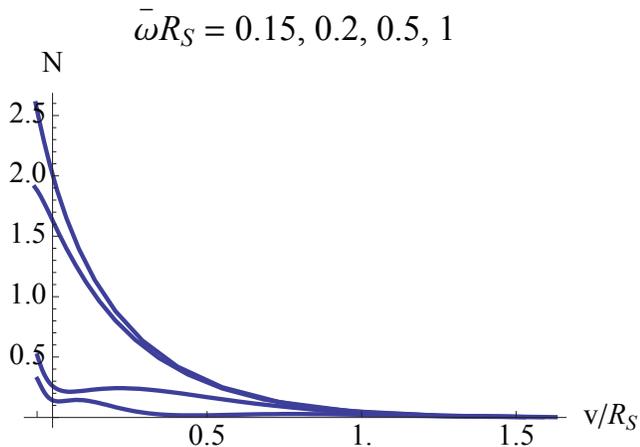}
\caption{Here we plot $N$ versus $v/R_s$ for various fixed values of $\bar{\omega} R_s$. The occupation number at all frequencies grows as the collapse progresses  (i.e. $v/R_s$ decreases) but never diverges. The occupation number would probably diverge when the singularity is hit at $v=-0.126$ because of the infinitely strong gravitational field, but our approximation breaks far from the Schwarzschild radius which is crossed at $v=0$, so we do not extend our plots all the way to $v=-0.126$. The curves are lower for higher values of $\bar{\omega} R_s$.}
\label{N_versus_v}
\end{figure}

In Fig.~\ref{N_versus_v} we plot $N$ versus $v/R_s$ for various fixed values of $\bar{\omega}R_s$. The frequency $\bar{\omega}$ is measured in time $v$. We can see that the occupation number at any frequency increases as $v/R_s$ decreases. Thus more particles are created as the shell reaches and crosses the horizon. However, the number of created particles does not diverge as $R(v) \rightarrow R_s$.

\begin{figure}[htbp]
\includegraphics{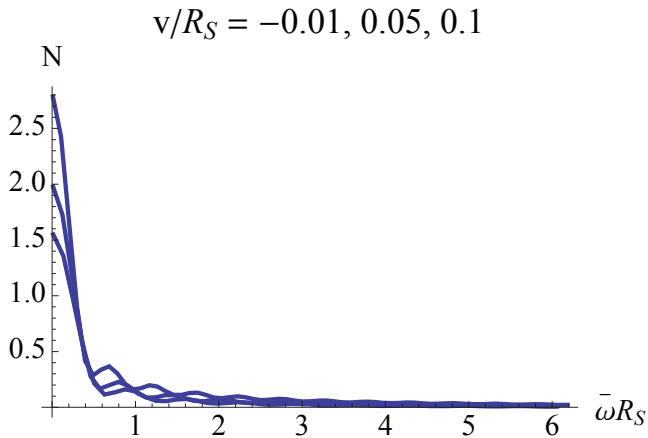}
\caption{Here we plot $N$ versus $\bar{\omega} R_s$ for various fixed values of $v/R_s$. The occupation number at any frequency grows as the collapse progresses  (i.e. $v/R_s$ decreases) but never diverges.}
\label{N_versus_w}
\end{figure}

We then numerically evaluate the spectrum of mode occupation numbers at any finite time and show the results in Fig.~\ref{N_versus_w} for several values of $v/R_s$. The first sign of non-thermality is the fact that the occupation number is non-divergent at $\bar{\omega}=0$, as opposed to the thermal Planck distribution in Eq.~(\ref{Planck Dist}).

For the values of parameters taken for plots in Fig.~\ref{N_versus_v} and Fig.~\ref{N_versus_w}, the Schwarzschild radius is crossed at $v=0$, while the singularity is reached at $v=-0.126$. However, we can not extend our plots significantly beyond $v=0$ since there our approximation breaks.

In Fig.~\ref{LnN_versus_w} we plot $\ln(1+1/N)$ versus $\bar{\omega}R_s$ for various values of $v/R_s$. As $v/R_s$ decreases (as the shell is collapsing), the curves decrease. A thermal spectrum should gives us a straight line, however, we see that is not the case here. The best one can do is to fit the low frequency part of the spectrum and get the temperature in that regime. We see that the temperature (of the low frequency part of the spectrum) practically remains constant near the Schwarzschild radius (near $v=0$). The numerical value that we get for the temperature is $T \approx 0.7/R_s$. Unlike the case of Schwarzschild coordinates, where the spectrum becomes thermal in the whole frequency range at late times, in Eddington-Finkelstein coordinates the spectrum never becomes thermal in the high frequency range.
We can not extend our quantitative analysis all the way to the singularity, since the approximation we used in Eq.~({\ref{app}) fails far from $R \approx R_s$.

We also note that all of the plots were made for the numerical value of the eigenvalue $y$ defined after Eq.~(\ref{schrod}) of $y=1$. Numerical experiments indicate that the spectrum and the temperature do not change significantly for different values of the eigenvalue $y$.

\begin{figure}[htbp]
\includegraphics{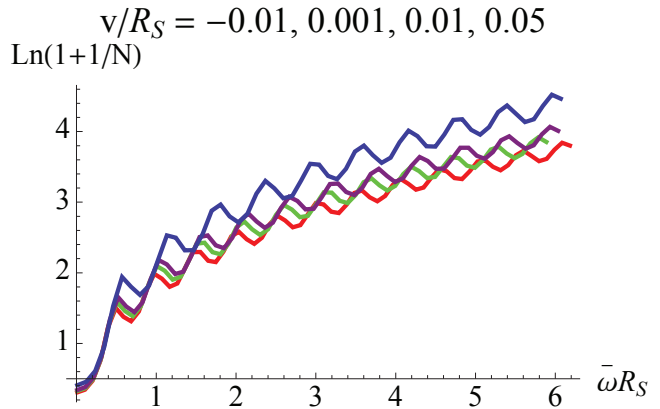}
\caption{Here we plot $\ln(1+1/N)$ versus $\bar{\omega} R_s$ for various fixed values of $v/R_s$. Clearly, one can not fit straight lines through these curves in the whole frequency regime. However, the low frequency part ($\bar{\omega}<R_s^{-1}$) does appear thermal with the temperature that is practically constant in time $v$. We get the numerical value of $T \approx 0.7/R_s$}
\label{LnN_versus_w}
\end{figure}

\section{Conclusions}

We studied the question of quantum radiation from a collapsing object from a point of view of an infalling observer. Precisely, we calculated the occupation number of particles whose frequencies are measured in the proper time of an infalling observer. As the collapsing object approaches its own Schwarzschild radius it excites fields that are propagating in this dynamical background. In the context of the functional Schrodinger formalism, the information about the excited radiation is contained in the time dependent wave function. The distribution function of the radiation depends on the wave function overlap between the initial state (vacuum) and the state at some later time. This formalism allows us to track the time evolution of the radiation distribution function. This was not possible in approximations usually used in similar setups, e.g. in Bogolyubov method where one matches the coefficients between the two static asymptotically flat spaces, i.e. Minkowski at the beginning and Schwarzschild at the end of the gravitational collapse.

We demonstrated several interesting results. Quantum radiation accompanies gravitational collapse since the metric describing the collapse is time dependent. However, the characteristics of the emitted radiation strongly depend on the space-time foliation of an observer. For the infalling observer in Schwarzschild coordinates the radiation distribution function is not quite thermal, though it becomes thermal when the collapsing object reaches its own Schwarzschild radius. We call such radiation Hawking-like or pre-Hawking radiation (as opposed to thermal Hawking radiation from a pre-existing horizon). In the absence of any backreaction, an asymptotic observer will observe a divergent flux of particles at infinity. However, the number of particles and their energies have different meaning for different observers. The radiation distribution function (i.e. the particle occupation number as a function of frequency) depends on the time coordinate that the observer is using. Naively, an infalling observer would need to encounter an infinite number of particles, but in finite amount of his proper time. However, it is only the occupation number of particles that is divergent, not the actual number of detected particles. An infalling observer can not detect particles whose wavelength is larger than his detector and therefore would miss most of them.

By fitting the approximate temperature corresponding to the radiation distribution function, we showed that the local temperature near the Schwarzschild radius, $R_s$, is large and diverges exactly at $R_s$.
This is expected since this local temperature has to be infinitely resifted in order to get a finite temperature of radiation as seen by the asymptotic observer. After all, an infalling observer in Schwarzschild coordinates is accelerated at any given moment, with acceleration which diverge exactly at $R_s$. Such an accelerated observer should ``see" divergent temperature at the horizon in agreement with the Unruh effect \cite{Unruh:1976db}.
This still does not mean that the horizon is a violent place for the infalling observer since local vacuum polarization energy density is usually thought to cancel out the divergence of the temperature energy density, which yields a finite total energy density..

The situation in Eddington-Finkelstein coordinates is quite different. This coordinate system is not singular at the horizon, and an infalling observer in these coordinates is not accelerated.
We find that the distribution of the particle occupation number is not thermal in the whole frequency range. By fitting the temperature only in the low frequency range ($\bar{\omega}<R_s^{-1}$) we find the temperature $T \approx 0.7/R_s$. This temperature is somewhat greater than the Hawking temperature as measured at infinity by an asymptotic observer $T_\infty = 1/(4\pi R_s)$,  which is in agreement with results obtained in different approaches (see e.g. \cite{Brynjolfsson:2008uc}). However, the spectrum never becomes thermal in the whole frequency range, so it is difficult to talk about the temperature as seen by such an observer.

\appendix

\section{Number of particles produced as a function of time}

We use the simple harmonic oscillator basis states but at a frequency $\bar{\omega}$ to keep track of different $\omega$'s in the calculation. To evaluate the occupation numbers at time $\tau<\tau_f$, we need only to set $\bar{\omega}=\omega(\tau_f)$, where the subscript ``f" stands for ``final". So
\begin{equation}
  \varphi(b)=\left(\frac{m\bar{\omega}}{\pi}\right)^{1/4}\frac{e^{-m\bar{\omega}b^2/2}}{\sqrt{2^nn!}}H_n(\sqrt{m\bar{\omega}}b)
\end{equation}
where $H_n$ are the Hermite polynomials. Then Eq.~(\ref{wavfunc}) together with Eq.~(\ref{c num}) gives
\begin{align}
  c_n=&\left(\frac{1}{\bar{\omega}\pi^2\rho^2}\right)^{1/4}\frac{e^{i\alpha}}{\sqrt{2^nn!}}\int d\zeta e^{-P\zeta^2/2}H_n(\zeta)\nonumber\\
           \equiv&\left(\frac{1}{\bar{\omega}\pi^2\rho^2}\right)^{1/4}\frac{e^{i\alpha}}{\sqrt{2^nn!}}I_n
\end{align}
where
\begin{equation}
  P=1-\frac{i}{\bar{\omega}}\left(\frac{\rho_\eta}{\rho}+\frac{i}{\rho^2}\right).
\end{equation}

To find $I_n$ consider the corresponding integral over the generating function for the Hermite polynomials
\begin{align}
  J(z)&=\int d\zeta e^{-P\zeta^2/2}e^{-z^2+2z\zeta}\nonumber\\
         &=\sqrt{\frac{2\pi}{P}}e^{-z^2(1-2/P)}
\end{align}
Since
\begin{equation}
  e^{-z^2+2z\zeta}=\sum_{n=0}^{\infty}\frac{z^n}{n!}H_n(\zeta)
\end{equation}
\begin{equation}
  \int d\zeta e^{-P\zeta^2/2}H_n(\zeta)=\frac{d^n}{dz^n}J(z)\Big{|}_{z=0}
\end{equation}
Therefore
\begin{equation}
  I_n=\sqrt{\frac{2\pi}{P}}\left(1-\frac{2}{P}\right)^{n/2}H_n(0).
\end{equation}
Since
\begin{equation}
  H_n(0)=(-1)^{n/2}\sqrt{2^nn!}\frac{(n-1)!!}{\sqrt{n!}}
\end{equation}
and $H_n(0)=0$ for odd $n$, we find the coefficient $c_n$ for even values of $n$,
\begin{equation}
  c_n=\frac{(-1)^{n/2}e^{i\alpha}}{(\bar{\omega}\rho^2)^{1/4}}\sqrt{\frac{2}{P}}\left(1-\frac{2}{P}\right)^{n/2}\frac{(n-1)!!}{\sqrt{n!}}.
\end{equation}
For odd $n$, $c_n=0$.

Next we find the number of particles produced. Let
\begin{equation}
  \xi=\left|1-\frac{2}{P}\right|.
\end{equation}
Then
\begin{align}
  N(\tau,\bar{\omega})&=\sum_{n=even}n\left|c_n\right|^2\nonumber\\
                                  &=\frac{2}{\sqrt{\bar{\omega}\rho^2}|P|}\xi\frac{d}{d\xi}\sum_{n=even}\frac{(n-1)!!}{n!!}\xi^n\nonumber\\
                                  &=\frac{2}{\sqrt{\bar{\omega}\rho^2}|P|}\xi\frac{d}{d\xi}\frac{1}{\sqrt{1-\xi^2}}\nonumber\\
                                  &=\frac{2}{\sqrt{\bar{\omega}\rho^2}|P|}\frac{\xi^2}{(1-\xi^2)^{3/2}}.
\end{align}
Inserting the expressions for $\xi$ and $P$, leads to
\begin{equation}
  N(\tau,\bar{\omega})=\frac{\bar{\omega}\rho^2}{\sqrt{2}}\left[\left(1-\frac{1}{\bar{\omega}\rho^2}\right)^2+\left(\frac{\rho_\eta}{\bar{\omega}\rho}\right)^2\right].
  \label{v occup}
\end{equation}

\section{Behavior of the function $\rho$ near the Schwarzschild radius}

To get an understanding of the number of particles created in the near horizon limit we need to investigate the behavior of the function $\rho$ in Schwarzschild coordinates near the Schwarzschild radius.

Near the horizon we can then write the velocity term as
\begin{equation}
 |R_{\tau}| \approx {\rm const } \equiv A.
  \label{hvelocity}
\end{equation}
In this limit the position of the shell is then, from Eq.~(\ref{R_tau}),
\begin{equation}
  R(\tau)\approx R_0-A\tau
\end{equation}
where $R_0$ is the initial position of the shell, so we can write
\begin{equation}
  \frac{\sqrt{1+R_{\tau}^2}}{|R_{\tau}|}\equiv C.
\end{equation}
Therefore the frequency becomes
\begin{equation} \label{frequency}
  \omega^2\approx\frac{\omega_0^2}{CB}.
\end{equation}
Therefore the auxiliary equation becomes
\begin{equation*}
  \rho_{\eta\eta}+\omega_0^2\frac{R_s}{C((R_0-R_s)-A\tau)}\rho=\frac{1}{\rho^3}
\end{equation*}
or using Eq.~(\ref{Eta}) we can write this as,
\begin{equation*}
  \frac{1}{C^2}\frac{d^2\rho}{d\eta^2}+\omega_0^2\frac{R_s}{C((R_0-R_s)-A\tau)}\rho=\frac{1}{\rho^3}.
\end{equation*}
after rescaling can be written as
\begin{equation}
  \frac{d^2f}{d\tau'^2}=-\frac{A^2\omega_0^{3/2}R_s^{3/4}C^{5/4}}{(R_0-R_s)^{11/4}}\left[\frac{f}{1-\tau'}-\frac{1}{f^3}\right]
  \label{F eq}
\end{equation}
where $\tau'=A\tau/(R_0-R_s)$, and $f=\sqrt{\omega_0}(R_s/C(R_0-R_s))^{1/4}\rho$. The boundary conditions are then
\begin{equation}
  f(0)=\left(\frac{R_s}{C(R_0-R_s)}\right)^{1/4}, \hspace{2mm} \frac{df(0)}{d\eta'}=0.
\end{equation}
The last term with the $1/f^3$ becomes singular as $f\rightarrow0$. Let us consider another equation with this term replaced by something more well behaved. For example consider,
\begin{equation}
  \frac{d^2g}{d\tau'^2}=-\frac{A^2\omega_0^{3/2}R_s^{3/4}C^{5/4}}{(R_0-R_s)^{11/4}}\left[\frac{g}{1-\tau'}-g\right]
  \label{G eq}
\end{equation}
with boundary conditions
\begin{equation}
  g(0)=\left(\frac{R_s}{C(R_0-R_s)}\right)^{1/4}, \hspace{2mm} \frac{dg(0)}{d\tau'}=0.
\end{equation}
Eq.~(\ref{G eq}) implies that $g(\tau')$ is monotonically decreasing as long as $g(\tau')>0$. Furthermore, it is decreasing faster than the solution for $f$ as long as $f<1$, since the $1/f^3$ term in Eq. (\ref{F eq}) is a larger ``repulsive" force than the $g$ term in Eq. (\ref{G eq}). Therefore we have
\begin{equation}
  f(\tau')\geq g(\tau')
\end{equation}
for all $\tau'$ such that $g(\tau')>0$.

The solution for $g$ is positive for all $\tau'$ and, in particular, $g(1)>0$ for all the values of $A^2\omega_0^{3/2}C^{5/4}R_s^{3/4}/(R_0-R_s)^{11/4}$ that we checked. Therefore $f(\tau')$ is positive, at least for a wide range.

Let $f_1=f(1)\neq0$. Then the equation for $f$ can be expanded near $\tau'=1$.
\begin{equation}
  \frac{d^2f}{d\tau'^2}=-\frac{A^2\omega_0^{3/2}R_s^{3/4}C^{5/4}}{(R_0-R_s)^{11/4}}\left[\frac{f_1}{1-\tau'}-\frac{1}{f_1^3}\right].
  \label{approx F}
\end{equation}
Integrating Eq.~(\ref{approx F}) we can then write
\begin{equation}
  \frac{df}{d\eta'}\sim\frac{A^2\omega_0^{3/2}R_s^{3/4}C^{5/4}}{(R_0-R_s)^{11/4}}f_1\ln(1-\tau')\rightarrow-\infty
\end{equation}
as $\tau'\rightarrow1$.
Hence $\rho(\tau=(R_0-R_s)/A)$ is strictly positive and finite while $\rho_{\tau}(\tau=(R_0-R_s)/A=-\infty$ for finite and non-zero $\omega_0$.

We are calculating the occupation number $N$ as a function of frequency $\omega$ at some fixed time.
From Eq.~(\ref{frequency}) we see that, in order to keep $\omega$ fixed in time, $\omega_0\rightarrow0$ as $B \rightarrow0$. Thus, $\omega$ varies with  $\omega_0$ and not with time.
Since $f=(R_s/C(R_0-R_s))^{1/4}$, and $f\rightarrow (R_s/C(R_0-R_s))^{1/4}$ for $\omega_0\rightarrow0$, we see that $\rho\rightarrow\infty$ and $\rho_{\tau}\rightarrow0$ as $\omega_0\rightarrow0$. This implies that the occupation number $N$ in Eq.~(\ref{N}) diverges as $\tau \rightarrow \tau_c$ since $B \rightarrow 0$ as $\tau \rightarrow \tau_c$.

\vskip.2cm {\bf Acknowledgment} \vskip.2cm
This paper grew out from comments that the anonymous referee made on the first version of the preprint.
The authors thank L. Thorlacius for very useful correspondence. DS acknowledges the financial support from NSF, award number PHY-0914893.


\end{document}